\documentclass[twocolumn,superscriptaddress,
 amsmath,amssymb,
 aps,prc
]{revtex4-2}

\usepackage{multirow,graphicx,url,braket}
\usepackage{dcolumn}
\usepackage{bm}
\usepackage{hyperref}
\usepackage[ulem=normalem]{changes}
\usepackage{threeparttable}
\begin{document}

\title{Nuclear size, electric monopole transitions, and the location of $0_2^+$ states}

\author{B. Maheshwari}
\email{bhoomika.physics@gmail.com}
\affiliation{Grand Accélérateur National d'Ions Lourds,
CEA/DSM-CNRS/IN2P3, Bvd Henri Becquerel, BP 55027, F-14076 Caen, France}

\author{K. Nomura}
\affiliation{Department of Physics, 
Hokkaido University, Sapporo 060-0810, Japan}
\affiliation{Nuclear Reaction Data Center, 
Hokkaido University, Sapporo 060-0810, Japan}

\author{P. Van Isacker}
\affiliation{Grand Accélérateur National d'Ions Lourds,
CEA/DSM-CNRS/IN2P3, Bvd Henri Becquerel, BP 55027, F-14076 Caen, France}

\date{\today}

\begin{abstract}

The work addresses the isotopic shift of nuclear radii for the even-even $^{36-52}$Ca isotopes using the interacting boson model (IBM) that includes the mixing from normal and intruder configurations. We obtain a good agreement between the calculated and experimental data, particularly for the dip at $^{48}$Ca. A direct correlation between nuclear size and electric monopole transitions is established to compute the electric monopole transition strengths, $\rho^2(E0)$. We further study the isotopic shift for the even-even $^{32-46}$Ar and $^{44-50}$Ti isotopes.   
\end{abstract}

\maketitle


\section{Introduction} 
\label{s_intro}
An important global property of an atomic nucleus is its size, that is,
its (charge) radius, which carries knowledge
about the nucleonic dynamics inside the nucleus due to the nuclear force.
Similar to electrons in an atom,
nucleons in a nucleus occupy discrete quantum levels
with shell gaps at nucleon numbers 2, 8, 20, 28, 50, 82, and 126
(colloquially known as ``magic" numbers).
At these nucleon numbers a nucleus experiences an increased stability
as compared to neighboring nuclei,
which can influence its charge radius.
The spin-orbit interaction plays a crucial role in the stability of nuclei
since it favors single-particle levels, denoted as $j_+$, with an orbital angular momentum $l$
that is aligned with the intrinsic spin $s$ of the nucleon.
For heavier nuclei, with mass numbers $A\gtrsim50$,
the effect of the spin-orbit interaction can be so strong
that a $j_+$ orbital with high $l$ intrudes
into a lower-lying shell of the harmonic oscillator.
This mechanism explains the magic numbers 28 and higher
that are observed in nuclei close to the stability line~\cite{mayer1955}.
Nevertheless, a fundamental understanding of the evolution of shell gaps
in neutron-rich or neutron-deficient nuclei, far removed from the stability line,
is still lacking~\cite{otsuka2020,ozawa2000,otsuka2001,sorlin2008,otsuka2005}.
In particular, their impact on charge radii remains a challenge
for modern experimental and theoretical nuclear physics.

Due to recent advances in radioactive-ion-beam production,
the charge radii of the calcium isotopes ($Z=20$)
are now known over a long range,
from proton-rich $^{36}$Ca~\cite{miller2019}
to neutron-rich $^{52}$Ca~\cite{ruiz2016}.
The measurements support the doubly-magic nature of $^{40}$Ca and $^{48}$Ca,
which have almost equal charge radii~\cite{ruiz2016}.
The local maximum of the charge radius of $^{44}$Ca
and its very steep increase beyond $^{48}$Ca
pose a serious challenge for nuclear theory.
If the ground states of even-even nuclei
can be described in terms of nucleonic pairs,
as is the basic assumption of the interacting boson model (IBM)~\cite{iachello87},
it is tempting to model the evolution of charge radii
with bosons~\cite{zerguine2008, zerguine2012, nomura2013, isacker2014}.

Weak shell gaps favor the formation
of deformed shapes~\cite{force2010,mizusaki2001,delaroche2007,CEA}
as well as the occurrence of shape coexistence~\cite{heyde2011,maheshwari2024,garrett2022,leoni2024}.
In the long chain of even-even $^{36-52}$Ca isotopes,
only one isotope, namely doubly-magic $^{40}$Ca,
has a first-excited state with angular momentum $J^\pi=0^+$,
which is a signature of shape coexistence.
This $0_2^+$ isomeric state has a half-life of $2.17$~ns~\cite{goro1966,schulz1967}
and is interpreted as a four-particle-four-hole excitation~\cite{ensdf}.
The $0_2^+$ state in another doubly-magic Ca isotope, $^{48}$Ca,
exists with a relatively shorter half-life of 0.22~ns~\cite{ensdf} with a $22.5\%$ of $E0$ branch.
The $0_2^+$ state in $^{42}$Ca is also known with a shorter half-life of 0.38~ns~\cite{ensdf} but with a weak $E0$ branch.
Electric monopole ($E0$) transitions from $0_2^+$ to the ground state
are known in the $^{40,42,44,48}$Ca isotopes~\cite{kibedi2022}
and their complementary $\rho^2 (E0)$ values follow an inverted parabola
with minima for $^{40,48}$Ca.
Although in most Ca isotopes an $E2$ transition
is possible to a lower-lying $2^+$ state,
the $E0$ behavior is possibly correlated with the $0_2^+$ half-lives especially when $B(E2; 0_2^+ \rightarrow 2_1^+ )$ is in the order of $\le$10 W.u.,
hinting at the importance of the $E0$ transition
to determine whether the $0_2^+$ state is isomeric or not.
$E0$ transitions connecting two $0^+$ states are special
since no photon emission is allowed
and the decay necessarily proceeds via electron conversion
and/or electron-positron pair emission~\cite{leoni2024}. 

In this article we aim to correlate the location of the $0_2^+$ state
with the size of the nucleus (i.e., its charge radius)
using a version of the IBM
that incorporates the mechanism of coexistence and configuration mixing,
referred hereafter as \mbox{IBM-CM}~\cite{duval1981}.
The puzzling behavior of the charge radii of the even-even $^{36-52}$Ca isotopes
so far has not been considered in the context of IBM,
to the best of our knowledge.
Electric-monopole transition strengths are also calculated in the same framework
and the double magicity of $^{40,48}$Ca is also explored~\cite{otsuka2001,otsuka2005}.
We further present similar results
for the even-even $^{32-48}$Ar and $^{44-50}$Ti isotopes
having two $sd$ proton holes and two $pf$ proton particles
with respect to Ca isotopes, respectively.

\section{Formalism}
\label{s_formalism}
The interacting boson model (IBM)~\cite{arima1975}
is comprised of $s$ and $d$ bosons with angular momenta $0^+$ and $2^+$,
representing monopole and quadrupole pairs of valence nucleons,
respectively~\cite{iachello87,OAI}.
The total number of bosons, denoted as $n_{\rm b}=n_s+ n_d$,
is determined by the microscopic interpretation
of active valence nucleons,
$n_{\rm b}=n_\pi+n_\nu$ where $n_\pi (n_\nu$)
equals the number of proton (neutron) particle or hole pairs
counted from  the nearest closed shell.
A distinctive feature of the IBM
is its ability to furnish the U(6) algebra
with generators formed by $s$ and $d_\mu$ ($\mu=0,\pm1,\pm2$) bosons.
The U(6) algebra is subsequently reduced to the U(5), SU(3), and SO(6) subalgebras,
providing vibrational, rotational and $\gamma$-unstable spectra, respectively.
The IBM Hamiltonian employed for phenomenological description
of low-energy states is expressed as follows:
\begin{eqnarray}
\hat H=\epsilon\,\hat n_d+a_1\hat L\cdot\hat L+a_2\hat Q\cdot\hat Q\;,
\label{eq:ibm}
\end{eqnarray}
where $\epsilon$ stands for the energy
of a single $d$ boson relative to that of an $s$ boson
and $\hat n_d=d^\dagger\cdot\tilde d$
denotes the $d$-boson number operator,
with $\tilde d_\mu=(-1)^\mu d_{-\mu}$.
The second term with strength $a_1$
represents the SO(3) Casimir operator,
with $\hat L=\sqrt{10}[d^\dagger\times\tilde d]^{(1)}$.
The third term with strength $a_2$
is the quadrupole-quadrupole interaction inducing quadrupolar deformation.
The corresponding operator reads
$\hat Q=s^\dagger\tilde d+d^\dagger\tilde s+\chi[d^\dagger\times\tilde d]^{(2)}$,
where $\chi$ is a parameter
which determines whether the nuclear shape
is prolate ($\chi<0$) or oblate ($\chi >0$)~\cite{iachello87}.

In the \mbox{IBM-CM} several independent Hamiltonians are introduced
for systems with $n_{\rm b},n_{\rm b}+2,n_{\rm b}+4,\ldots$ bosons,
associated with 0p--0h, 2p--2h, 4p--4h, $\ldots$ particle-hole excitations, respectively.
The unperturbed eigenstates of the separate Hamiltonians
are allowed to mix in order to incorporate configuration mixing~\cite{duval1981}.
Here we consider two configurations, i.e.,
the normal or regular $[n_{\rm b}]$ and the intruder $[n_{\rm b}+2]$ configuration.
The full \mbox{IBM-CM} Hamiltonian reads
\begin{eqnarray}
\hat H'=\hat H_{\rm reg}+(\hat H_{\rm int}+\Delta)+\hat{V}_{\rm mix}\; ,
\label{eq:hamiltonian}
\end{eqnarray}
where $\hat H_{\rm reg}$ and $\hat H_{\rm int}$ represent the Hamiltonians
for the regular (0p-0h) and intruder (2p-2h) boson spaces, respectively,
and each has the form given by Eq.~(\ref{eq:ibm}).
$\Delta$ represents the energy needed
to promote a nucleon pair from a major oscillator shell to the next.
The last term in Eq.~(\ref{eq:hamiltonian}) is the interaction
which admixes the different boson spaces and which is given by
\begin{eqnarray}
\hat V_{\rm mix}=
\alpha(s^\dagger s^\dagger+\tilde s\tilde s)+
\beta[d^\dagger\times d^\dagger+\tilde d\times\tilde d]^{(0)}\;,
\label{mixibm}
\end{eqnarray}
where $\alpha$ and $\beta$ parametrize the mixing strength.
In all calculations presented here
fixed values for $\alpha$ and $\beta$ are taken for each isotopic chain,
which in addition are assumed to be equal, $\alpha=\beta$. $^{48}$Ca is an exception to this assumption.
The Hamiltonian~(\ref{eq:hamiltonian}) is diagonalized
in the full $[n_{\rm b}]\oplus[n_{\rm b}+2]$ space
to obtain excitation energies and wave functions~\cite{isacker}.
A common core of $^{40}$Ca is used for the \mbox{IBM-CM} calculations.

The question arises whether a version of the IBM should be used that respects isospin invariance. Specifically, in IBM-3~\cite{elliott1980} a neutron-proton boson with isospin T=1 is introduced besides the usual neutron-neutron and proton-proton bosons of IBM-2, and in IBM-4~\cite{elliott1981} one considers additionally a neutron-proton boson with T=0. Such more elaborate versions of the IBM are important in nuclei with valence neutrons and protons in the same orbitals since it allows the definition of the isospin quantum number for a boson system. As a result, these extended boson models have a more solid microscopic connection with the shell model. However, so far no version of IBM-3 or IBM-4 has been developed that incorporates both regular and intruder configurations. Also, a procedure is known to project from IBM-3 to IBM-1~\cite{lac1995}, and in the latter system the mixing of several configurations is well established.

The procedure of fitting the \mbox{IBM-CM} Hamiltonian parameters
is described in Ref.~\cite{maheshwari2024}
and all parameters used for the present analysis are listed in Table~\ref{tab:ibmcm}.
The calculated level energies are compared with the experimental data,
wherever available, in Table~\ref{tab:level}.
An overall agreement between the experimental data and calculated results
has been achieved by fixing $\epsilon$ based on the location of the $2_1^+$ state,
and $\Delta$ based on that of the $0_2^+$ state.
The energies of other levels in the spectra
are then determined from the $a_1$ and $a_2$ parameters. 
The spectra of most of the fitted nuclei
are compatible with $\chi=0$ in both the $[n_{\rm b}]$ and $[n_{\rm b}+2]$ spaces
with the exception of the $^{44,50}$Ca, $^{32,34,36}$Ar, and $^{44,46,48}$Ti isotopes.
The results for the $^{42,44,46}$Ca isotopes
have already been discussed in Ref.~\cite{maheshwari2024}. In $^{50}$Ca, a prolate deformed SU(3) $[n_{\rm b}+2]$ space
(with $\chi =-1.33$) is assumed.
In $^{32,34}$Ar an oblate deformed $\overline{\rm{SU(3)}}$ $[n_{\rm b}]$ space
(with $\chi =+1.33)$
together with a prolate deformed SU(3) $[n_{\rm b}+2]$ space
(with $\chi =-1.33$) is assumed.
In $^{36}$Ar an oblate deformed $\overline{\rm{SU(3)}}$ $[n_{\rm b}]$ space
(with $\chi =+1.33)$ is assumed with a vibrational U(5) $[n_{\rm b}+2]$ space.
In the stable $^{38}$Ar isotope and onwards
$\chi=0$ is found to be sufficient in both $[n_{\rm b}]$ and $[n_{\rm b}+2]$ spaces.
This is also consistent with the shape description of these lighter Ar isotopes in the literature,
see the recent review~\cite{leoni2024}.
In $^{44,46,48}$Ti a $[n_{\rm b}]$ space is assumed with $\chi =-1$. 
It may be noted that the fitting of parameters is primarily based on the energies of few lower-lying levels. Such parameters may likely to fall insufficient to describe the other observables such as $B(E2)$s.   

In the IBM the charge radius operator is taken as~\cite{iachello87,zerguine2008,zerguine2012}
\begin{eqnarray}
\hat T(r^2)=\langle r^2\rangle_{\rm c}+
\kappa_{\rm reg}\hat n_{\rm b}+
\eta_{\rm reg}\frac{\hat n_d}{\hat n_{\rm b}}\;,
\label{eq:radii}
\end{eqnarray}
where the first term is the square of the charge radius of the core nucleus.
The second term refers to the increase of the charge radius
due to the addition of two nucleons.
The third term describes the contribution arising from the deformation.
The same expression is valid for the intruder $[n_{\rm b}+2]$ configuration
with the parameters $\kappa_{\rm int}$ and $\eta_{\rm int}$
for the second and third terms, respectively.
In the standard IBM the increase of the charge radius
due to the addition of two nucleons is assumed to be linear
as it is taken proportional to the number of bosons.
Note, however, that in order to ensure this linear behavior
at the middle of the valence shell
a change of sign of the parameter $\kappa_{\rm reg}$ is needed
since the interpretation of the bosons
changes from a pair of particles to a pair of holes~\cite{iachello87}. In IBM-CM, this linear behavior breaks down especially when one shifts from $[n_b]$ space to the $[n_b+2]$ space. To maintain in the \mbox{IBM-CM} a linear increase of the second term in Eq.~(\ref{eq:radii})
with the number of valence nucleon pairs,
it is necessary to impose the relation
$\kappa_{\rm int}=\frac{n_{\rm b}}{n_{\rm b}+2}\kappa_{\rm reg}$~\cite{harder1997}.
One calculates, therefore, the isotopic shift of the charge radius
with respect to the core as follows:
\begin{eqnarray}
\delta\langle r^2\rangle^{A}&=&
\langle\hat T(r^2)\rangle_{0_1^+}^A-\langle r^2\rangle_{\rm c}
\nonumber\\
&=&2\kappa_{\rm reg}n_{\rm b}
+\frac{\eta_{\rm reg}}{n_{\rm b}}\langle0_1^+ |\hat n_d|0_1^+\rangle^A_{n_{\rm b}}
\nonumber\\ 
&&+\frac{\eta_{\rm int}}{n_{\rm b}+2}\langle0_1^+|\hat n_d| 0_1^+\rangle^A_{n_{\rm b}+2}\;.
\label{eq:is}
\end{eqnarray}

\begin{table*}[!htb]
\caption{\label{tab:ibmcm}
Parameters of the \mbox{IBM-CM} Hamiltonian for the nuclei considered in this study. }
\centering
\begin{ruledtabular}
\resizebox{0.95\textwidth}{!}{
\begin{tabular}{cccccccccccccccc}
\multirow{2}{*}{Nucleus} & 
\multicolumn{2}{c}{$\epsilon$ (MeV)} &
\multicolumn{2}{c}{$a_1$ (MeV)} &
\multicolumn{2}{c}{$a_2$ (MeV)} &
\multicolumn{2}{c}{$\chi$} &
\multirow{2}{*}{$\alpha=\beta$ (MeV)} & 
\multirow{2}{*}{$\Delta$ (MeV)}  \\
\cline{2-3}
\cline{4-5}
\cline{6-7}
\cline{8-9}
 & 
$[n_{\rm b}]$ & $[n_{\rm b}+2]$ &
$[n_{\rm b}]$ & $[n_{\rm b}+2]$ &
$[n_{\rm b}]$ & $[n_{\rm b}+2]$ &
$[n_{\rm b}]$ & $[n_{\rm b}+2]$ &
 & &
\\
\hline
$^{36}$Ca &  1.1 & 1.3 & 0.32 & 0.10 & -0.05 & -0.05 & 0.0 & 0.0 & 0.25 & 1.5 \\
$^{38}$Ca &  0.9 & 0.7 & 0.18 & 0.02 & -0.05 & -0.03 & 0.0 & 0.0 & 0.25 & 1.5 \\
$^{42}$Ca &   0.4 & 0.4 & 0.20 & -0.01 & -0.05 & -0.05 & 0.0 & 0.0 & 0.25 & 1.0 \\
$^{44}$Ca &  0.6 &  0.4 & 0.05 &  -0.004 & -0.05 &  -0.05 & 0.0 &  -1.0 & 0.25 & 1.1 \\
$^{46}$Ca &  0.8 &  0.8 & 0.18 &  -0.04 & -0.05 &  -0.05 & 0.0 &  0.0 & 0.25 & 1.0 \\
$^{48}$Ca &   2.1 & 2.2 & 0.60 & -0.14 & -0.035 & -0.06 & 0.0 & 0.0 & 0.30 & 2.0 \\
$^{50}$Ca &   0.8 & 1.95 & 0.09 & 0.10 & -0.05 & -0.15 & 0.0 & -1.33 & 0.25 & 0.65 \\
\hline
$^{32}$Ar &  1.35 & 0.95 & 0.18 & 0.05 & -0.125 & -0.08 & 1.33 & -1.33 & 0.15 & 1.55 \\
$^{34}$Ar &   1.35 & 0.95 & 0.18 & 0.05 & -0.125 & -0.08 & 1.33 & -1.33 & 0.15 & 1.55  \\
$^{36}$Ar &   1.25 & 1.15 & 0.15 & -0.06 & -0.138 & -0.065 & 1.33 & 0.0 & 0.15 & 1.55  \\
$^{38}$Ar &   1.35 & 0.9 & 0.18 & 0.015 & 0.045 & 0.025 & 0.0 & 0.0 & 0.15 & 1.55  \\
$^{40}$Ar &   1.35 & 0.35 & 0.033 & 0.001 & 0.065 & 0.0 & 0.0 & 0.0 & 0.15 & 1.55  \\
$^{42}$Ar &  0.95 & 0.9 & 0.025 & 0.005 & 0.038 & 0.025 & 0.0 & 0.0 & 0.15 & 1.55 \\
$^{44}$Ar &  0.95 & 0.9 & 0.015 & 0.005 & 0.04 & 0.025 & 0.0 & 0.0 & 0.15 & 1.55  \\
$^{46}$Ar &  1.15 & 0.9 & 0.075 & 0.005 & 0.0 & 0.005 & 0.0 & 0.0 & 0.15 & 1.55 \\
\hline
$^{44}$Ti &  0.7 & 0.5 & 0.018 & 0.014 & -0.05 & -0.05 & -1.0& 0.0 & 0.30 & 0.90 \\
$^{46}$Ti &   0.6 & 0.5 & 0.02 & -0.02 & -0.12 & -0.095 & -1.0 & 0.0 & 0.30 & 0.90  \\
$^{48}$Ti &  0.6 & 0.5 & 0.013 & -0.018 & -0.18 & -0.16 & -1.0 & 0.0 & 0.30 & 0.90 \\
$^{50}$Ti &  1.65 & 1.45 & -0.06 & 0.08  & 0.01 & 0.06 & 0.0 & 0.0 & 0.30 & 0.90 \\  
\end{tabular}}
\end{ruledtabular}
\end{table*}
 
In this framework $E0$ transitions can be directly correlated
with the nuclear size, that is, with the charge radius of the ground state.
The electric monopole transition strength $\rho(E0)$
between the second-excited $0_2^+$ and ground $0_1^+$ states
in the $[n_{\rm b}] \oplus [n_{\rm b}+2]$ space is given by 
\begin{eqnarray}
\rho(E0;0_2^+\rightarrow0_1^+)&=&
\frac{2en_{\rm b}}{eR^2}
\Bigg(\frac{\eta_{\rm reg}}{n_{\rm b}}
\langle0_1^+|\hat n_d|0_2^+\rangle^A_{n_{\rm b}}
\nonumber\\
&&+\frac{\eta_{\rm int}}{n_{\rm b}+2}
\langle0_1^+|\hat n_d|0_2^+\rangle^A_{n_{\rm b}+2}\Bigg)\;,
\label{eq:rho2}
\end{eqnarray}
where $R=1.02A^{1/3}$~fm is the radius of an ideal-spherical nucleus
and the coefficient $1.02$ is chosen for the lighter-mass nuclei based on the charge radius of $^{40}$Ca~\cite{angeli2013}.
The number of active valence nucleons
appears as $2n_{\rm b}$ in the numerator of Eq.~(\ref{eq:rho2}).
The first and second terms arise due to the contributions from the deformation
in the respective $[n_{\rm b}]$ and $[n_{\rm b}+2]$ spaces.
The $E0$ strengths are usually expressed in terms of the square of Eq.~(\ref{eq:rho2}), $\rho^2(E0)$,
which is a dimensionless quantity.
The same parameters $\eta_{\rm reg}$ and $\eta_{\rm int}$,
as obtained from the isotopic shifts by the application of Eq.~(\ref{eq:is}),
are used to calculate $E0$ strengths,
implying therefore a direct correlation between charge radii and electric monopole transitions.

\begin{table*}[!htb]
\caption{\label{tab:level}
Experimental energies~\cite{nndc} and energies calculated in the \mbox{IBM-CM}
of some yrast and yrare levels in the nuclei considered in this study. }
\centering
\begin{ruledtabular}
\resizebox{0.95\textwidth}{!}{
\begin{tabular}{cccccccccccccc}
\multirow{2}{*}{Nucleus} & 
\multicolumn{6}{c}{Experimental E(MeV)} &
\multicolumn{6}{c}{Calculated E(MeV)}  \\
\cline{2-3}
\cline{4-5}
\cline{6-7}
\cline{8-9}
\cline{10-11}
\cline{12-13}
 & 
${0_1^+}$ & ${0_2^+}$ &
$2_1^+$ & $2_2^+$ &
$4_1^+$ & $4_2^+$
& 
${0_1^+}$ & ${0_2^+}$ &
$2_1^+$ & $2_2^+$ &
$4_1^+$ & $4_2^+$\\
\hline
$^{36}$Ca &  0.0 & 2.830 & 3.045 & 4.710 &  &  & 0.0 & 2.853 & 3.163 & 4.737 & 8.102 & 9.424 \\
$^{38}$Ca &  0.0 & 3.083 & 2.213 & 3.683 & 5.816$^*$ &  & 0.0 & 3.074 & 2.202 & 3.863 & 4.883 & 5.823 \\
$^{40}$Ca &  0.0 & 3.352 & 3.904 & 5.248 & 5.278 & 6.507 & \multicolumn{6}{c}{------core------} \\
$^{42}$Ca &  0.0 & 1.837 & 1.524 & 2.424 & 2.752 & 3.253 & 0.0 & 1.951 & 1.637 & 2.463 & 2.596 & 3.396  \\
$^{44}$Ca &  0.0 & 1.883 & 1.157 & 2.656 & 2.283 & 3.044 & 0.0 & 1.808 & 1.041 & 2.053 & 2.309 & 3.204\\
$^{46}$Ca &  0.0 & 2.423 & 1.346 & 3.022 & 2.574 & 3.859 & 0.0 & 2.304 & 1.577 & 2.777 & 2.644 & 3.810\\
$^{48}$Ca &  0.0 & 4.283 & 3.831 & 6.105$^*$ & 4.503  & & 0.0 & 4.320 & 3.792 & 6.184 & 4.638 & 7.075 \\
$^{50}$Ca &  0.0 & 4.475$^*$ & 1.026 & 3.002$^*$ & 4.515$^*$ & & 0.0 & 4.301 & 1.225 & 6.143 & 3.996 & 9.399 \\
\hline
$^{32}$Ar &  0.0 & & 1.867 & & & &  0.0 & 2.525 & 1.914 & 3.008 & 4.372 & 5.990 \\
$^{34}$Ar &  0.0 & 3.873 & 2.091 & 3.287 & 4.631$^\dagger$ & & 0.0 & 2.595 & 2.160 & 3.253 & 4.668 & 6.469  \\
$^{36}$Ar &  0.0 &  & 1.970 & 4.440 & 4.414 & 6.136 & 0.0 & 2.958 & 2.192 & 4.022 & 4.439 & 6.057  \\
$^{38}$Ar &  0.0 & 3.376 & 2.167 & 3.936 & 5.349 & 6.053$^*$ & 0.0 & 3.312 & 2.260 & 4.294 & 5.298 & 5.998 \\
$^{40}$Ar &  0.0 & 2.120 & 1.460 & 2.524 & 2.892 & 3.515 & 0.0 & 2.239 & 1.451 & 2.427 & 2.831 & 3.404 \\
$^{42}$Ar &  0.0 &  & 1.208 & 2.485 & 2.413$^*$ & 3.096 & 0.0 & 1.908 & 1.165 & 2.024 & 2.370 & 3.040 \\
$^{44}$Ar &  0.0 & 0.750? & 1.158 & 1.610$^*$ & 2.746$^*$ & & 0.0 & 2.109 & 1.194 & 2.148 & 2.354 & 3.141   \\
$^{46}$Ar &  0.0 & & 1.570 & 3.489 & 3.892 & & 0.0 & 2.400 & 1.629 & 2.834 & 3.793 & 5.004 \\
\hline
$^{44}$Ti & 0.0 & 1.904 & 1.083 & 2.530 & 2.454 & 3.364 & 0.0 & 2.024 & 1.041 & 2.203 & 2.400 & 3.562 \\
$^{46}$Ti & 0.0 & 2.611 & 0.889 & 2.961 & 2.009 & 3.848$^*$ & 0.0 & 2.525 & 0.886 & 2.251 & 2.108 & 3.570  \\
$^{48}$Ti & 0.0 &  2.997   & 0.983 & 2.421 & 2.295 &  3.239 & 0.0 & 3.092 & 0.889 & 2.410 & 2.186 & 3.963  \\
$^{50}$Ti &  0.0 & 3.868 & 1.553 & 4.309 & 2.674 & 5.186 & 0.0 & 3.692 & 1.659 & 3.466 & 2.716 & 4.412 \\ 
\end{tabular}}
\begin{tablenotes}
        \raggedright
        \item $^*:$ Spin-parity assignment is not confirmed.
        \item $?:$ Based on NNDC evaluators~\cite{nndc}.
        \item $^\dagger:$ A. R. L. Kennington \textit{et al.}~\cite{kennington2021}
\end{tablenotes}
\end{ruledtabular}
\end{table*}

To complement the discussion on the location of $0_2^+$ state, also results of the NSM are presented.
All calculations are performed with the KSHELL code~\cite{kshell}.
The center-of-mass correction is taken care of as default. 
The results for $^{36,38}$Ca are obtained 
with the SDPF-MU interaction~\cite{utsuno2012}
with a full $sd$ ($sdpf$) valence space for the protons (neutrons). 
The NSM calculations for the even-even $^{42-52}$Ca isotopes
are performed in the $pf$ valence space with the KB3G interaction~\cite{poves2001}.
The NSM calculations for $^{32-36}$Ar
are performed with the SDPF-MU interaction
in the $sd$ $(sdpf)$ valence space for protons (neutrons)
while the $sd$ ($pf$) valence space is used for protons (neutrons) in $^{40-46}$Ar.
This valence space is usually found to be sufficient for these Ar isotopes,
especially for the yrast spectra~\cite{speidel2008}.
In $^{44-50}$Ti the NSM calculations are performed with the KB3G interaction
in the $pf$ valence space for both protons and neutrons.

\begin{figure}[!htb]
\centering
\includegraphics[width=0.47\textwidth,height=0.25\textwidth]{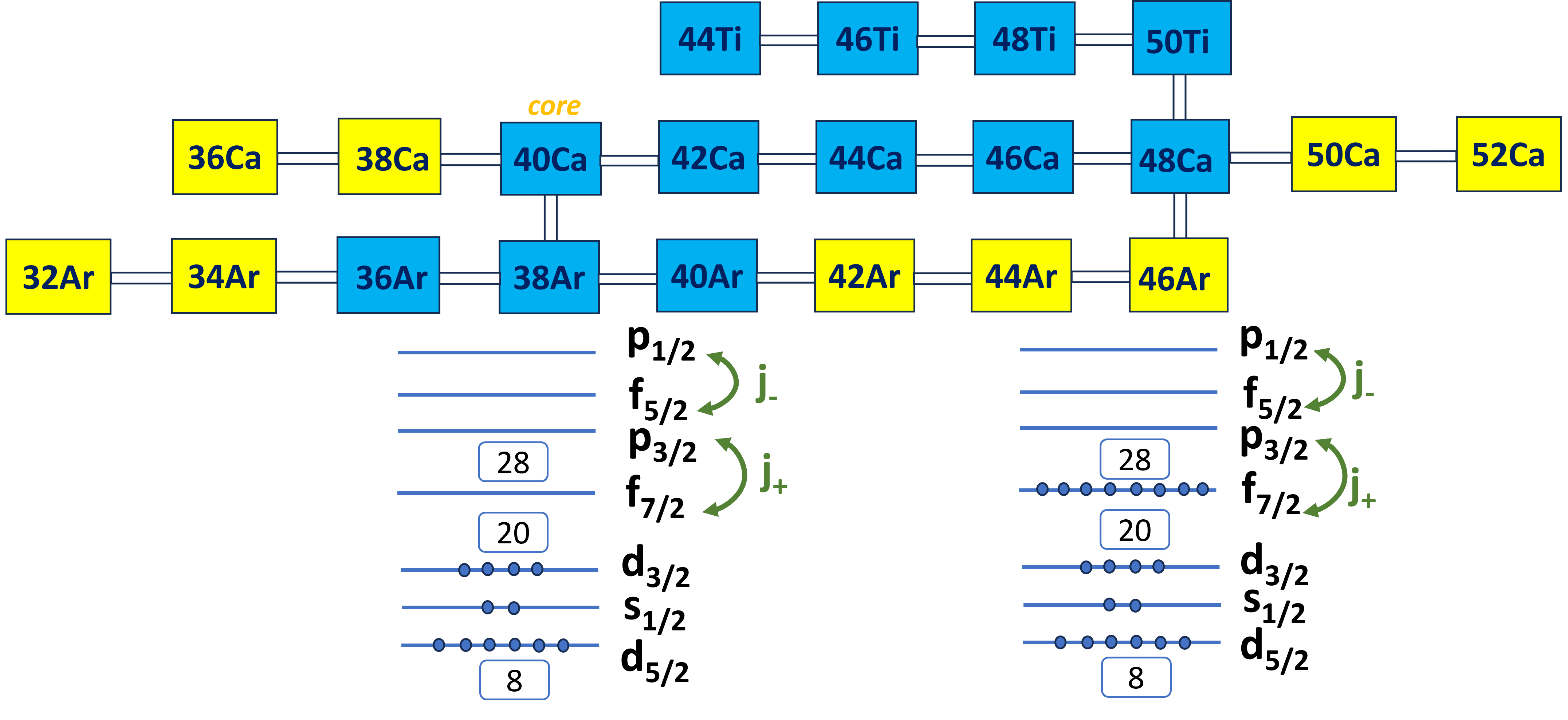}
\caption{(Color online)
Schematic picture of the region of interest which includes nuclei with a known charge radius~\cite{wholfahrt1981,anastassov1994,blaum2008,ruiz2016,miller2019}.
Stable isotopes with natural abundances are in blue
and others are in yellow~\cite{mass, nndc}.
Active single-particle orbitals are also shown together with the shell gaps.
This includes $N=28$, the first magic gap due to the spin-orbit interaction. The two level schemes are just to show the $Z,$ or $N=20$ and 28 closed shell configurations. The vertical double lines connecting nuclei is depicted for closed shells, while the horizontal ones simply refer to a given isotopic chain.}
\label{fig:region}
\end{figure}

\begin{figure}[!htb]
\centering
\includegraphics[width=0.49\textwidth]{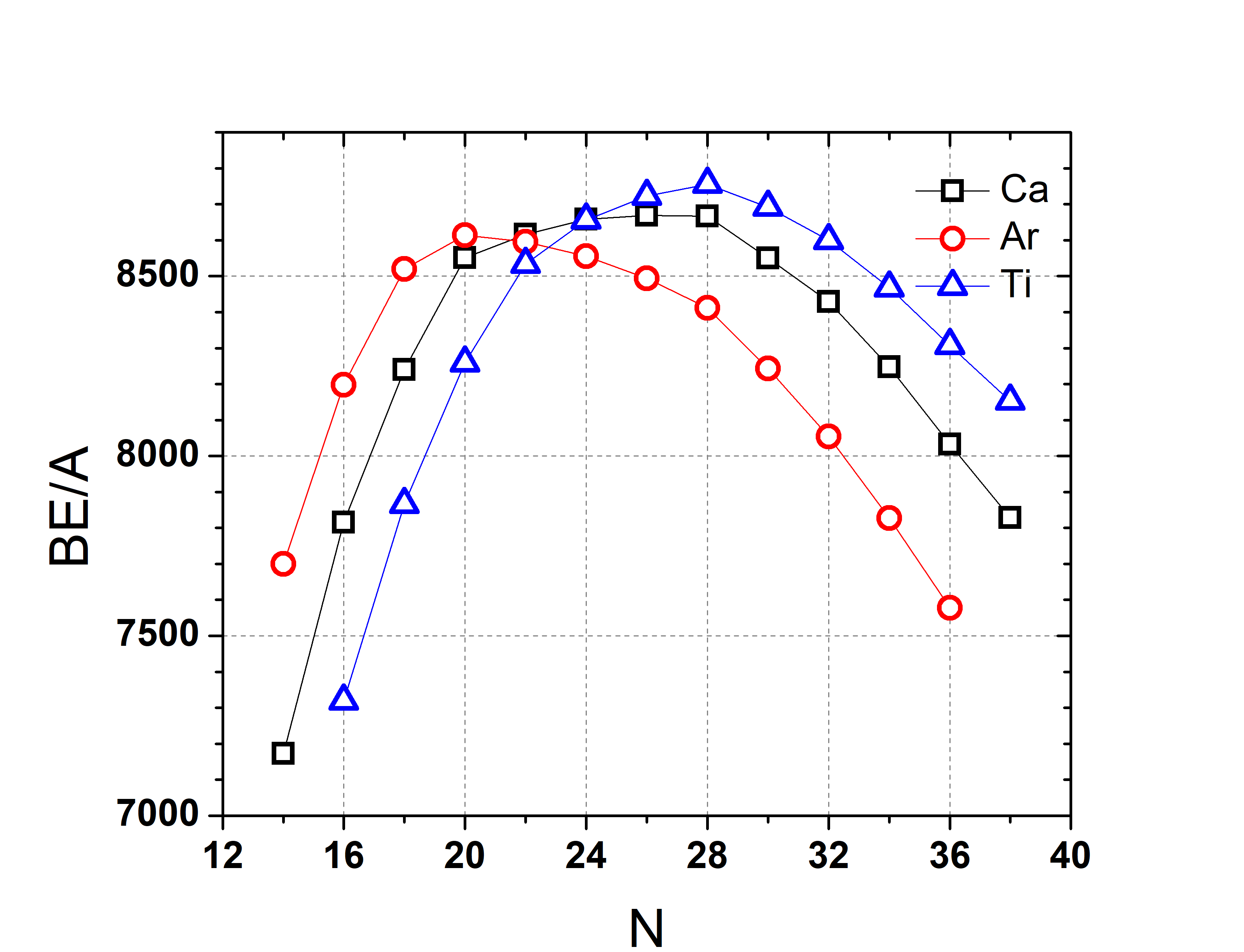}
\caption{(Color online)
Empirical binding energy per nucleon (BE/$A$) for even-even Ca, Ar, and Ti isotopes~\cite{nndc}. }
\label{fig:be}\end{figure}

\begin{figure}[!htb]
\centering
\includegraphics[width=0.45\textwidth]{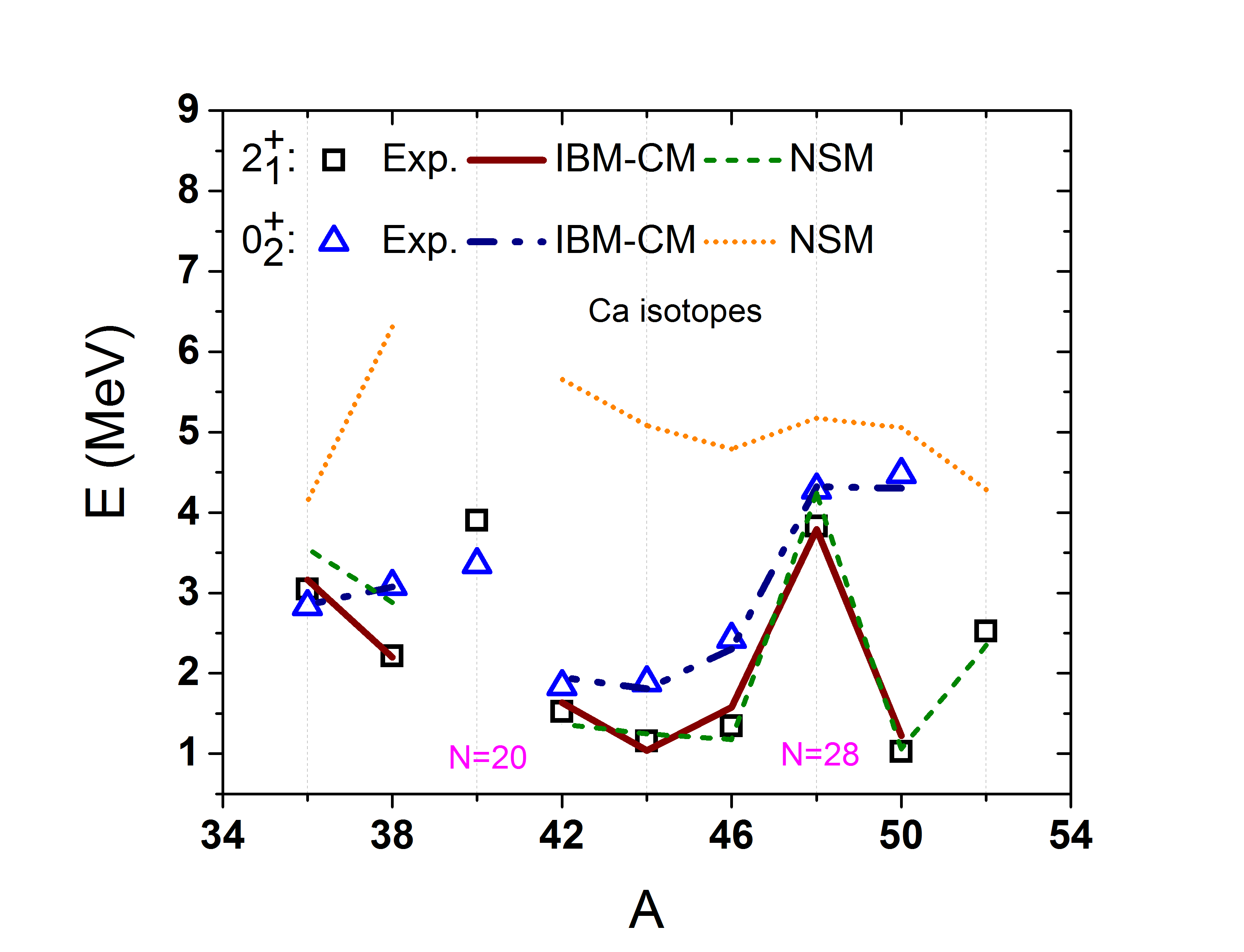}
\caption{(Color online)
Experimental~\cite{ensdf} and calculated energies $E$
of the yrast $2_1^+$ and yrare $0_2^+$ levels in the even-even $^{36-52}$Ca isotopes.}
\label{fig:caenergy}
\end{figure}

\section{Discussion} 
\label{s_discussion}
Figure~\ref{fig:region} refers to the region of interest in and around the $Z=20$ shell closure,
from even-even Ca ($Z=20$) to Ar ($Z=18$) and Ti ($Z=22$) isotopes,
for which data on charge radii
are available~\cite{wholfahrt1981,anastassov1994,blaum2008,ruiz2016,miller2019}.
Both $^{40}$Ca ($Z=20,N=20$) and $^{48}$Ca ($Z=20,N=28$) are (almost) stable,
consistent with their doubly-magic character,
while in the Ar and Ti isotopes
the semi-magic $^{38}$Ar ($Z=18,N=20$) and $^{50}$Ti ($Z=22,N=28$) are stable
but $^{46}$Ar ($Z=18,N=28$) and $^{42}$Ti ($Z=22,N=20$) are not.  

Figure~\ref{fig:be} shows the empirical binding energy per nucleon (BE/$A$)
for even-even Ca, Ar, and Ti isotopes~\cite{nndc}.
The binding energy per nucleon is rather flat for the Ca isotopes
between doubly-magic $^{40}$Ca and $^{48}$Ca,
and it steeply drops off outside this region.
In contrast, in the Ar and Ti isotopes,
the semi-magic $^{38}$Ar and $^{50}$Ti have maximal binding energy per nucleon
while $^{46}$Ar and $^{42}$Ti,
although conceivably also semi-magic with $N=28$ and $N=20$, respectively,
have a much lower BE/$A$.
The question thus arises whether this characteristic behavior of BE/$A$
is reflected in the evolution of the charge radii.

\subsection{The calcium isotopes}
\label{ss_ca}
We first apply the \mbox{IBM-CM} to the even-even $^{36-52}$Ca isotopes
with parameters as listed in Table~\ref{tab:ibmcm}.
The doubly-magic nucleus $^{40}$Ca is assumed to be the inert core
and hence the boson numbers for the regular configuration
in the even-even $^{42-52}$Ca isotopes are $1,2,3,\dots,6$, respectively.
An overall good agreement between the experimental and calculated low-energy levels
is achieved, as listed in Table~\ref{tab:level}.
The predicted and experimental energies of the $2_1^+$ and $0_2^+$ levels
are compared in Fig.~\ref{fig:caenergy}.
The present calculation,
which includes mixing of regular and intruder configurations,
agrees nicely with the experimental energies of the $2_1^+$ and $0_2^+$ levels in the Ca isotopes. In $^{52}$Ca, the experimental data are quite limited to fix the IBM-CM analysis which is therefore not performed.
Also, the measured level energy of $2_1^+$ in $^{50}$Ca
displays a sharp dip at 1.026~MeV,
to be compared with the surrounding isotopes,
namely the doubly-magic $^{48}$Ca ($2_1^+$ at 3.831~MeV)
and $^{52}$Ca ($2_1^+$ at 2.563 MeV).
This suggests to measure low-energy levels in $^{52}$Ca,
especially since this nuclide is getting attention
as regards a potential $N=32$ sub-shell closure~\cite{sorlin2008,ruiz2016}. 

In a recent experimental work on the structure of $^{36}$Ca~\cite{lalanne2022}
the importance of core excitations was pointed out,
especially for the second-excited $0_2^+$ state.
The present \mbox{IBM-CM} analysis supports this analysis.
Notably, the $0_2^+$ levels in $^{36,38}$Ca
are at almost equal energies (see Table~\ref{tab:level})
whereas the $2_1^+$ level in $^{36}$Ca is at higher energy than in $^{38}$Ca.
The latter may be related to a possible $N=16$ shell closure~\cite{sorlin2008}.
In $^{38}$Ca the yrast $2_1^+$ state
is most likely dominated by the $d_{3/2}^2$ configuration,
similar to $^{50}$Ca where it is dominated by $p_{3/2}^2$,
both involving $j=3/2$.
Similarly, the $0_2^+$ levels in $^{36,38}$Ca are at similar energies
while the $2_1^+$ excitation energies are vastly different.
The $2_1^+$ energy is indicative of shell gaps
whereas that of $0_2^+$ is sensitive to possible core excitations.
Shell gaps, when identified with effective single-particle energies,
are known to evolve with nucleon number 
but they may also state-dependent in a single nuclide
due to a self-organization mechanism~\cite{otsuka2019}.
The structure of the $2_1^+$ and $0_2^+$ states
is therefore crucial for understanding the evolution of shell gaps in nuclei.         

It may be recalled that the NSM interactions usually explain the variation of the excitation energy of the $2_1^+$ level but overestimates the location of $0_2^+$ states in pf-shell nuclei, such as the case in the $^{42,44,46}$Ca isotopes~\cite{maheshwari2024},
indicating the importance of core excitations for the yrare states in Ca. The present set of NSM calculations in Fig.~\ref{fig:caenergy} is reliable for yrast states only.

\begin{figure}[!htb]
\centering
\includegraphics[width=0.45\textwidth]{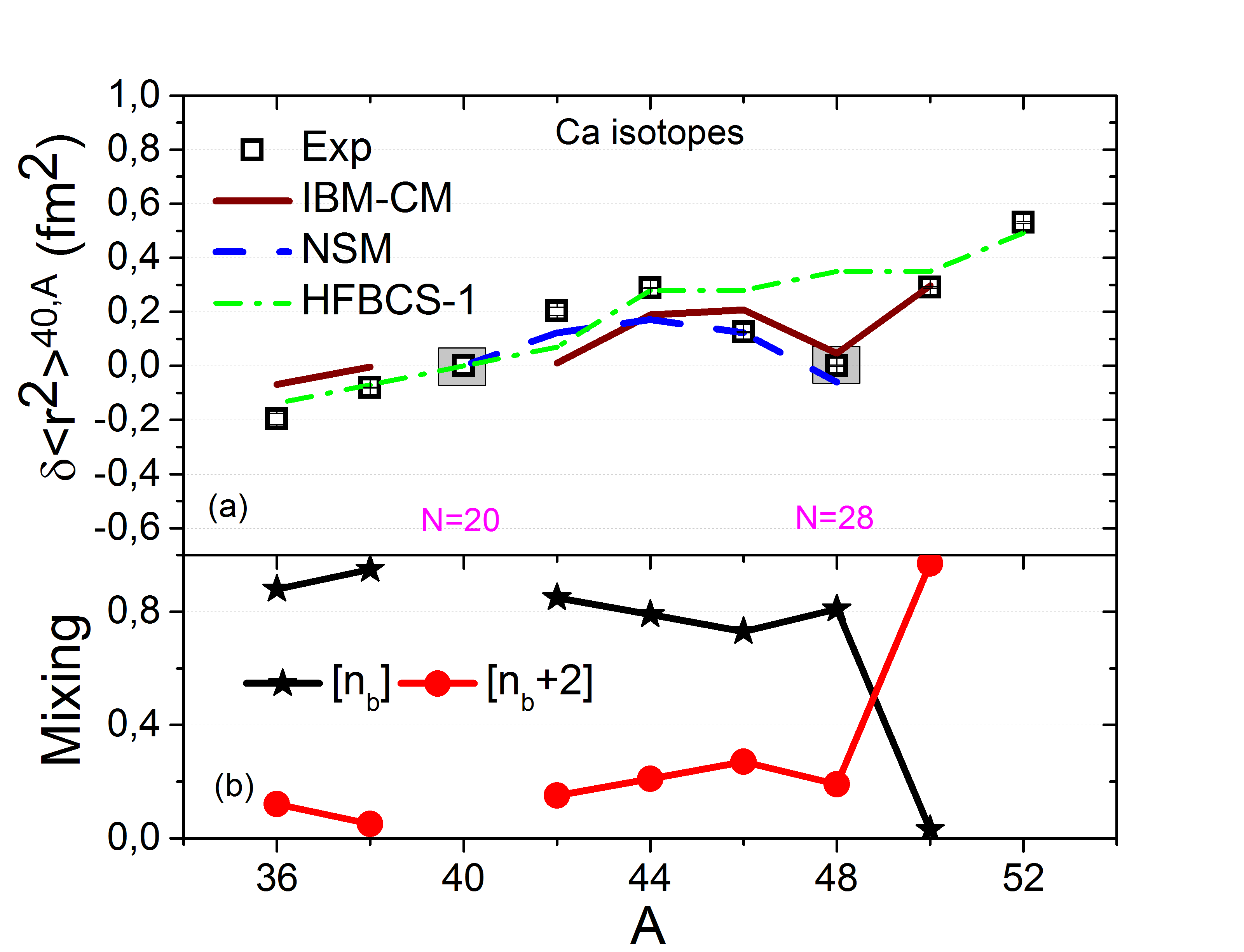}
\caption{(Color online)
(a) Experimental~\cite{ruiz2016,miller2019} and calculated isotopic shifts
in the even-even $^{36-52}$Ca isotopes.
Results of \mbox{IBM-CM}, NSM~\cite{caurier2001}, and HFBCS-1~\cite{goriely2001} calculations are shown.
The stable isotopes and the doubly-magic $^{40,48}$Ca are shown in grey. The error bars on data are within the size of symbol.
(b) The mixing of the $[n_{\rm b}]$ and $[n_{\rm b}+2]$ spaces
for the ground states of the even-even $^{36-52}$Ca isotopes.}
\label{fig:Ca}
\end{figure}

\begin{figure}[!htb]
\centering
\includegraphics[width=0.45\textwidth]{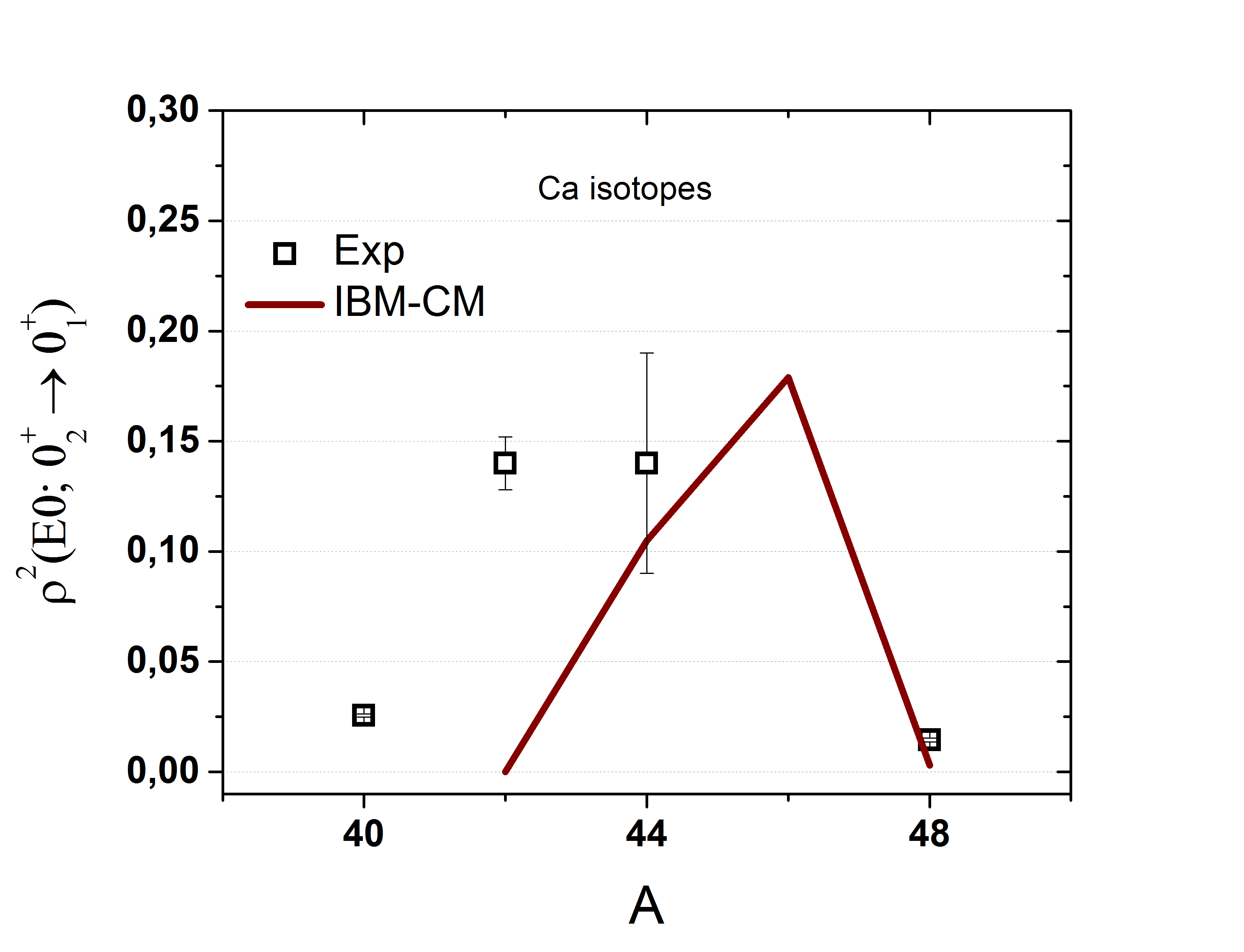}
\caption{(Color online)
Experimental~\cite{kibedi2022} and calculated $\rho^2(E0; 0_2^+ \rightarrow 0_1^+)$ values
within the \mbox{IBM-CM} in the even-even $^{36-52}$Ca isotopes.}
\label{fig:carho}
\end{figure}

With wave functions fixed from level energies,
one can compute the charge radii using Eq.~(\ref{eq:radii}),
and the corresponding isotopic shift (of the charge radii) using Eq.~(\ref{eq:is}).
The parameter $\kappa_{\rm reg}=0.0005$~fm$^2$
is adjusted to the overall slope of the isotopic shift for the even-even $^{42-52}$Ca isotopes
but otherwise the main features are controlled by two parameters,
$\eta_{\rm reg}=8.69$~fm$^2$ and $\eta_{\rm int}=0.59$~fm$^2$.
The absolute values of these parameters are then used
to compute the isotopic shifts of $^{36,38}$Ca
but taken to be negative since in these isotopes bosons correspond to hole pairs.
The corresponding \mbox{IBM-CM} isotopic shifts
are compared with the data in the even-even $^{36-52}$Ca isotopes in Fig.~\ref{fig:Ca}(a).
The calculated results are in an overall agreement
with the data for the long chain of Ca isotopes.
The \mbox{IBM-CM} does not explain the isotopic shift of $^{38,42}$Ca
due to limited number of active bosons,
which influences the contributions of $\hat n_d$ matrix elements.
However, the dip for $^{48}$Ca and the sharp increase at $^{50}$Ca
is well reproduced due to the change in the dominant wave function,
as depicted in Fig.~\ref{fig:Ca}(b).

\begin{table}[!htb]
\caption{\label{tab:mixing}
The mixing probabilities of the $[n_{\rm b}]$ and $[n_{\rm b}+2]$ spaces calculated in \mbox{IBM-CM}
for the $2_1^+$ and $0_2^+$ states.}
\centering
\begin{ruledtabular}
\resizebox{0.45\textwidth}{!}{
\begin{tabular}{ccccc}
\multirow{2}{*}{Nucleus} & 
\multicolumn{2}{c}{$2_1^+$} &
\multicolumn{2}{c}{$0_2^+$}  \\
\cline{2-3}
\cline{4-5}
 & 
$[n_{\rm b}]$ & $[n_{\rm b}+2]$ &
$[n_{\rm b}]$ & $[n_{\rm b}+2]$ \\
\hline
$^{36}$Ca & 0.83 & 0.17 & 0.14 & 0.86  \\
$^{38}$Ca &  0.93 & 0.07 & 0.05 & 0.95 \\
$^{40}$Ca &  \multicolumn{4}{c}{-----core-----}\\
$^{42}$Ca &  0.52 & 0.48 & 0.15 & 0.85  \\
$^{44}$Ca &  0.70 & 0.30 & 0.75 & 0.25 \\
$^{46}$Ca &  0.45 & 0.55 & 0.81 & 0.19 \\
$^{48}$Ca &  0.22 & 0.78 & 0.25 & 0.75 \\
$^{50}$Ca &  0.02 & 0.98 & 0.10 & 0.90 \\
\hline
$^{32}$Ar &  0.86 & 0.14 & 0.14 & 0.86 \\
$^{34}$Ar &  0.85 & 0.15 & 0.15 & 0.85 \\
$^{36}$Ar &  0.95 & 0.05 & 0.97 & 0.03 \\
$^{38}$Ar &  0.99 & 0.01 & 0.01 & 0.99 \\
$^{40}$Ar &  0.93 & 0.07 & 0.91 & 0.09\\
$^{42}$Ar &  0.97 & 0.03 & 0.98 & 0.02 \\
$^{44}$Ar &  0.96 & 0.04 & 0.97 & 0.03\\
$^{46}$Ar &  0.92 & 0.08 & 0.95 & 0.05\\
\hline
$^{44}$Ti & 0.72 & 0.28 & 0.81 & 0.19\\
$^{46}$Ti & 0.56 & 0.44 & 0.75 & 0.25 \\
$^{48}$Ti & 0.34 & 0.66 & 0.64 & 0.36  \\
$^{50}$Ti & 0.91 & 0.09 & 0.85 & 0.15  \\ 
\end{tabular}}
\end{ruledtabular}
\end{table}

There is a blocking effect for the core excitations as one reaches $^{48}$Ca.
It may be related to the blocking of excitations from the $d$ to $f$ orbital
as neutrons are fully filling the $f_{7/2}$ at $N=28$~\cite{ricci1969}.
This also contributes to the stability of $^{48}$Ca, and to a smaller nuclear size.
The states in nuclei in the middle of $f_{7/2}$, $^{44,46}$Ca,
become more collective from the mixing
between the regular and intruder spaces of the \mbox{IBM-CM},
as shown in Fig.~\ref{fig:Ca}(b).
This involves a change in the neutron-neutron interaction
in going from $^{42}$Ca, involving mainly the $d_{3/2}$ and $f_{7/2}$ orbitals,
to $^{48}$Ca, involving mainly the $f_{7/2}$ and $p_{3/2}$ orbitals.
Below $^{48}$Ca, the coupling between the $d$--$f$ orbitals,
which share the same principal quantum number, results in a more attractive force,
causing a contraction of the charge radius~\cite{otsuka2001}.
In contrast, above $^{48}$Ca the coupling between the $f$--$p$ orbitals,
which have different principal quantum numbers,
leads to a larger charge radius.
This is why the kink in the charge radii around $^{40}$Ca
is softer than the one around $^{48}$Ca,
which is therefore explained by the fact that $N=28$
is the first magic gap due to the spin-orbit interaction.
This also explains why the $0_2^+$ state
has a longer half-life in $^{40}$Ca than it has in $^{48}$Ca.
Figure~\ref{fig:Ca}(b) presents the mixing
of the $[n_{\rm b}]$ and $[n_{\rm b}+2]$ spaces in the \mbox{IBM-CM}
for the ground states of the even-even $^{36-52}$Ca isotopes.
In $^{50,52}$Ca the major component of the ground state is in the $[n_{\rm b}+2]$ space
which leads to a larger nuclear size, as is observed.
For completeness the $[n_{\rm b}]$ and $[n_{\rm b}+2]$ probabilities
for the $2_1^+$ and $0_2^+$ states are listed in Table~\ref{tab:mixing}.
The mixing of two spaces is clearly visible
but it is not always the case that the $0_2^+$ state
is dominated by $[n_{\rm b}+2]$ space.    

We also show in Fig.~\ref{fig:Ca} the NSM results adopted from Caurier \textit{et al.}~\cite{caurier2001} for the isotopic shifts
of the even-even $^{40-48}$Ca isotopes. Such NSM estimates explain the dip in charge radii at $^{48}$Ca using the number of protons lifted from $sd-$ to the $pf-$ shell. A similar exercise with available $sdpf$ interactions using the prescription of ref.~\cite{caurier2001} is not suitable, since these effective interactions were mainly designed for neutron-rich $sd-$shell nuclei and can not reproduce the location of $0_2^+$ state, as depicted for $^{36,38}$Ca isotopes in Fig.~\ref{fig:caenergy}. This already hints at a possible connection between the charge radius and the location of $0_2^+$ state, since the interaction used in ref.~\cite{caurier2001} explains the overall variation of charge radii by correcting the original interaction~\cite{retamosa1997} for the agreement of $0_2^+$ state by subtracting the double counting of pairing renormalization. 

In shell model picture, in the ground state of $^{52}$Ca the $f_{7/2}$ and $p_{3/2}$ orbitals
are nearly fully occupied, causing a subshell gap at $N=32$.
This does not imply that the $^{52}$Ca charge radius
is the same as those of $^{40,48}$Ca
since the BE/$A$ also affects the nuclear size,
as evidenced by the data in Fig.~\ref{fig:be}.
The large fragmentation of the wave functions,
covering the active neutron $f_{5/2}, p_{3/2}, p_{1/2}$ orbitals
separate from the first intruder $f_{7/2}$ orbital,
is the possible reason behind the large value of the isotopic shift in $^{52}$Ca.
This may be attributed to the location
of the spin-orbit partners $f_{7/2}$ and $f_{5/2}$~\cite{poves2001}:
a higher spin-orbit splitting leads to a larger charge radius.
This mechanism is also known in $^{208}$Pb region
where the intruder orbital and its spin-orbit partner, $i_{13/2}$ and $i_{11/2}$,
have many orbitals lying in between them~\cite{bhuyan2021}.
Also, the $N/Z$ ratio is almost the same in $^{52}$Ca and $^{212}$Pb. 

For comparison, we also show in Fig.~\ref{fig:Ca} the isotopic shift
taken from the Hartree-Fock nuclear mass table, as HFBCS-1~\cite{goriely2001}. 
This approach reproduces the overall evolution of charge radii
in the even-even $^{36-52}$Ca isotopes,
including the large value at $^{52}$Ca,
but not the dip in the isotopic shift at $N=28$.

We use the same parameters $\eta_{\rm reg}$ and $\eta_{\rm int}$
to calculate the $\rho^2(E0;0_2^+\rightarrow0_1^+)$ values
in the even-even $^{40-48}$Ca isotopes.
The results are compared with the experimental data~\cite{kibedi2022},
where known,  in Fig.~\ref{fig:carho}.
No calculated result is shown for $^{40}$Ca,
as it is taken as the inert core.
In $^{42}$Ca the calculated $\rho^2(E0;0_2^+\rightarrow0_1^+)$ value deviates from the data,
as was the case for its isotopic shift.
In $^{44,48}$Ca, on the other hand, the \mbox{IBM-CM} results
are in reasonable agreement with the data.
This suggests the existence of a correlation
between the charge radii of ground states
and the energy of $0_2^+$ levels,
which in turn influences the $0_2^+\rightarrow0_1^+$ $E0$ transitions.
The $\rho^2(E0;0^+_2\rightarrow0^+_1)$ values
are very small in both $^{40}$Ca and $^{48}$Ca,
hinting at a possible connection between the location of $0_2^+$ states and evolution of shell gaps.

\begin{figure}
\centering
\includegraphics[width=0.45\textwidth]{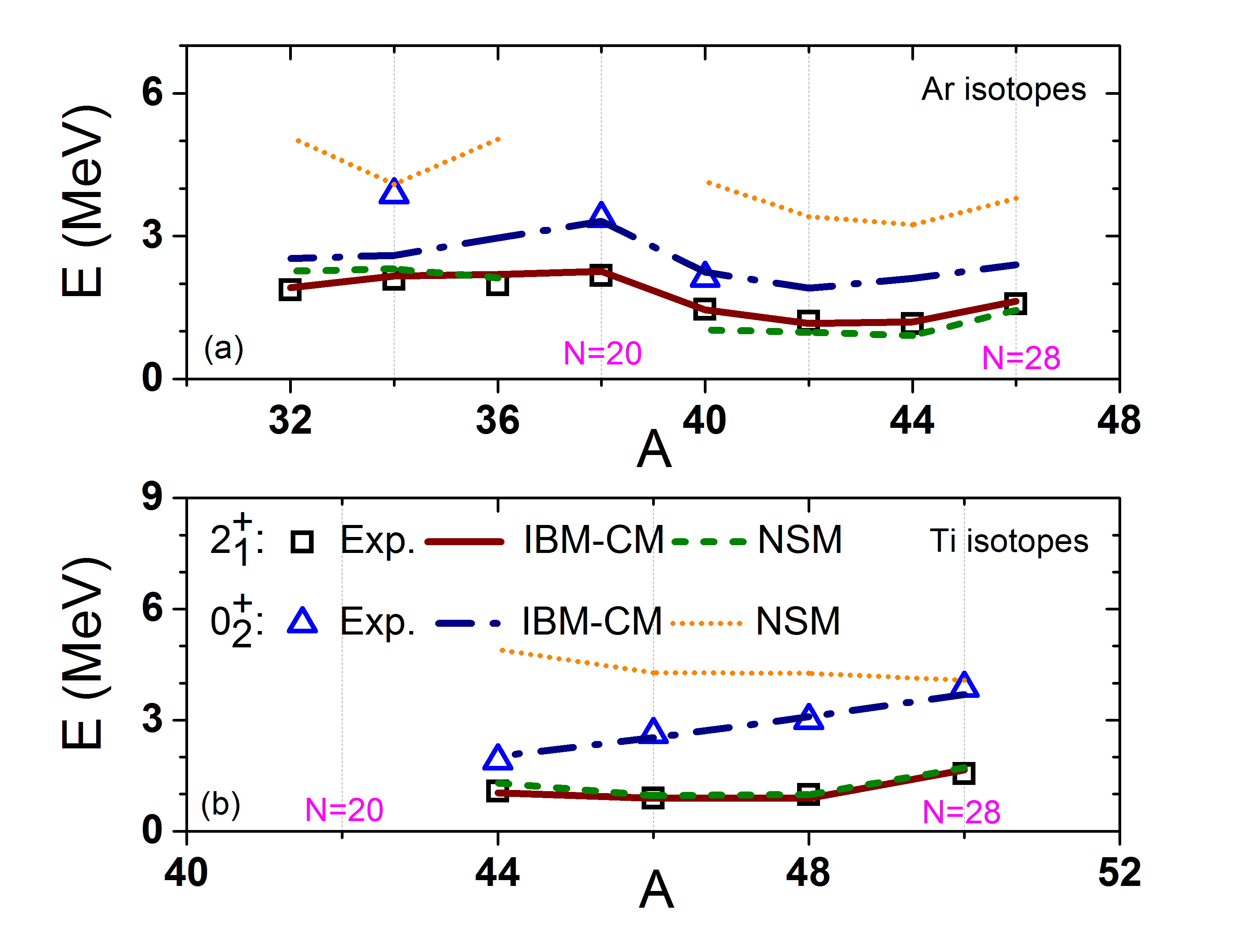}
\caption{(Color online)
Experimental~\cite{ensdf} and calculated energies $E$
of the yrast $2_1^+$ and yrare $0_2^+$ levels in even-even
(a) $^{32-46}$Ar and (b) $^{44-50}$Ti isotopes.}
\label{fig:artienergy}
\end{figure}

%

\begin{figure}
\centering
\includegraphics[width=0.48\textwidth]{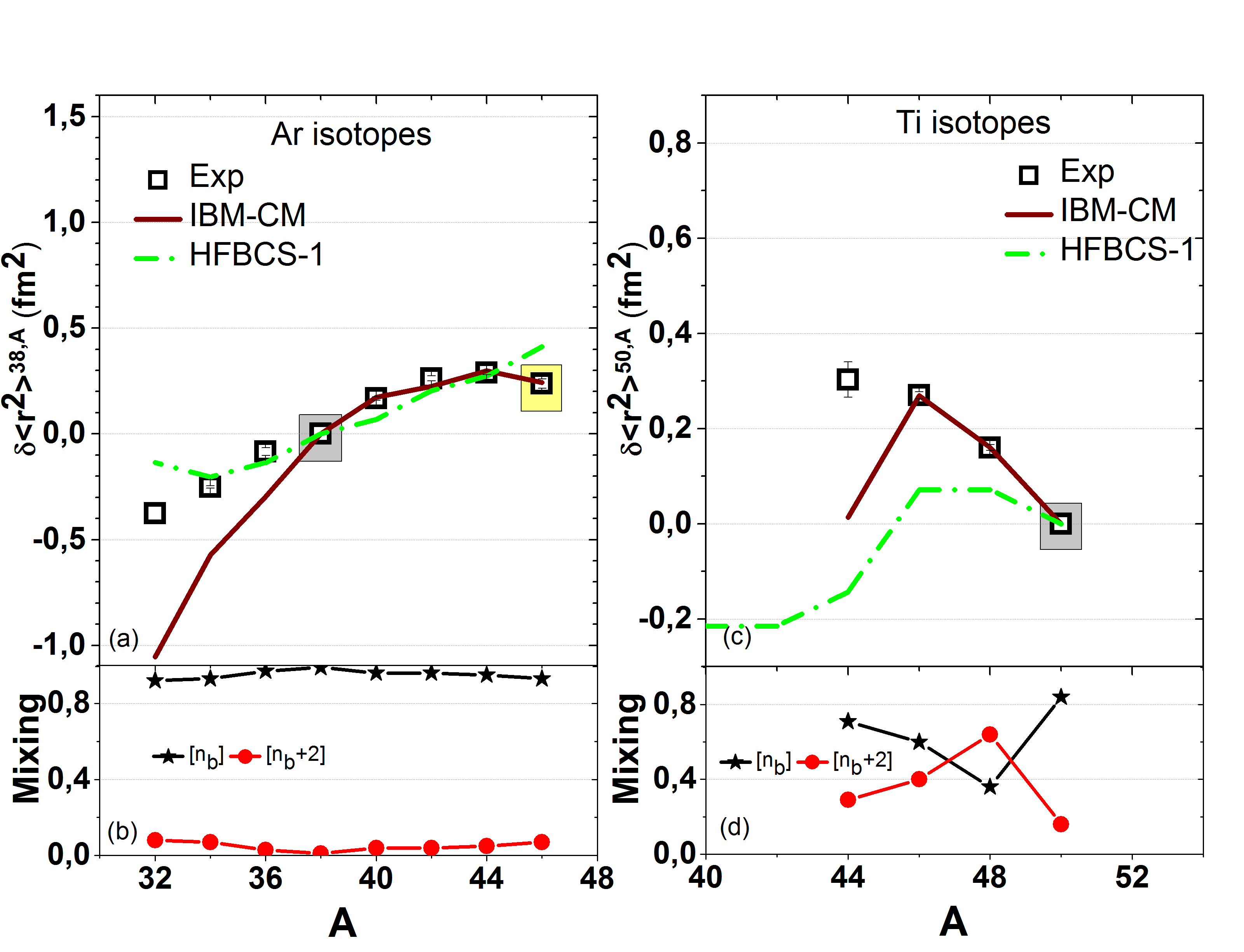}
\caption{(Color online)
(a) Experimental~\cite{wholfahrt1981,anastassov1994,blaum2008} and calculated isotopic shifts
in the even-even $^{32-46}$Ar isotopes.
Results of \mbox{IBM-CM}, and HFBCS-1~\cite{goriely2001} calculations are shown.
The stable semi-magic isotope is indicated in grey while the unstable semi-magic isotope is in yellow.
(b) The mixing of the $[n_{\rm b}]$ and $[n_{\rm b}+2]$ spaces
for the ground states of the even-even $^{32-46}$Ar isotopes.
(c) and (d) are same as (a) and (b) but for the even-even $^{44-50}$Ti isotopes.}
\label{fig:arti}
\end{figure}

\subsection{The argon and titanium isotopes}
\label{ss_arti}
We further perform \mbox{IBM-CM} calculations
for the even-even $^{32-46}$Ar and $^{44-50}$Ti isotopes
using the parameters as listed in Table~\ref{tab:ibmcm}.
$^{40}$Ca is taken to be core.
The excitation energies of the $2_1^+$ and $0_2^+$ levels
calculated with the \mbox{IBM-CM}
are compared with the experimental data in Fig.~\ref{fig:artienergy}.
In the Ar isotopes the \mbox{IBM-CM} results agree very well with the data
except for $^{34}$Ar where the predicted $0_2^+$
is much lower in energy than observed as shown in Fig.~\ref{fig:artienergy}(a). The measurement of $0_2^+$ state in neighboring $^{32,36}$Ar isotopes may help in future analysis.
The spectroscopy in neutron-deficient $^{32-36}$Ar isotopes is very limited,
which hinders a satisfactory IBM analysis. 
In the Ti isotopes the agreement between the \mbox{IBM-CM} results and the data
is quite nice for all the considered isotopes in Fig.~\ref{fig:artienergy}(b).
We also compare with NSM energies for both the Ar and Ti isotopes in Fig.~\ref{fig:artienergy}.
The NSM results in both chains explain the yrast $2_1^+$ state
while overestimating the energy of the yrare $0_2^+$ level
except in $^{34}$Ar and $^{50}$Ti.


With bosonic wave functions fixed from level energies,
we compute the isotopic shifts in the even-even $^{40-46}$Ar isotopes in Fig.~\ref{fig:arti}(a),
with fixed fitted values for $\kappa_{\rm reg}=0.024$~fm$^2$,
$\eta_{\rm reg}=5.47$~fm$^2$,
and $\eta_{\rm int}=0.01$~fm$^2$.
These parameters are then used to estimate the isotopic shifts
in the even-even $^{32-38}$Ar isotopes,
which were further normalized with respect to $^{38}$Ar to compare with data.
The mixing of the $[n_{\rm b}]$ and $[n_{\rm b}+2]$ spaces in the \mbox{IBM-CM}
is shown in Fig.~\ref{fig:arti}(b) for the ground states of the even-even $^{32-46}$Ar isotopes.
A very weak mixing appears to be reasonable in all the considered Ar isotopes
with a relatively pure configuration in the semi-magic stable nucleus $^{38}$Ar. The $0_2^+$ state in $^{38}$Ar mainly arises due to the core-excitations as was expected in the $E0$ measurement conducted in 1975~\cite{souw1975}, though IBM-CM analysis could not explain the corresponding order of $\rho^2(E0)$. The very large value of $\chi$ parameter required to reproduce the level energies in $^{32,34,36}$Ar eventually influences the $\hat{n}_d$ matrix elements in both the regular and intruder IBM configuration spaces, leading to a larger deformation effects controlling the charge radii if compared to neighboring Ca isotopes.  

A similar \mbox{IBM-CM} study is made for the even-even $^{44-50}$Ti isotopes,
the results of which are shown in Fig.~\ref{fig:arti}(c)
with $\kappa_{\rm reg}=-0.0002$~fm$^2$,
$\eta_{\rm reg}=4.1$~fm$^2$,
and $\eta_{\rm int}=-1.5$~fm$^2$.
The calculation could not explain the charge radius of $^{44}$Ti.
The mixing of $[n_{\rm b}]$ and $[n_{\rm b}+2]$ spaces in the ground states of these Ti isotopes
is depicted in Fig.~\ref{fig:arti}(d).
 A relatively pure $[n_{\rm b}]$ space is found in semi-magic $^{50}$Ti with $\chi$ parameter being nearly zero. In $^{44,46,48}$Ti isotopes, a large value of $\chi$ parameter for explaining the level energies, eventually influences the deformation effects arising from $\hat{n}_d$ matrix elements in IBM-CM analysis and consequently, changing the charge radii. The Table~\ref{tab:mixing} lists the mixing probabilities obtained in the IBM-CM analysis for the first $2_1^+$ and excited $0_2^+$ states in these Ti isotopes. It may be noted that the mixing supported by the first $2_1^+$ state for $^{46,48}$Ti in their bosonic wave functions is in line with the measured $g-$factor values~\cite{ernst2000}, suggesting them to be of collective character.

The NSM results on the charge radii for both the Ar and Ti isotopic chains are not known in literature, and could not be estimated in absence of the interaction by Caurier \textit{et al.}~\cite{caurier2001}. The semi-magic $^{38}$Ar is clearly more stable to the next semi-magic $^{46}$Ar, as also shown in Fig.~\ref{fig:arti}(a). The difference in BE/$A$ explains qualitatively why the charge radii in the two semi-magic Ar isotopes, $^{38}$Ar and $^{46}$Ar, are quite different, in contrast to the $^{40,48}$Ca isotopes, which have similar BE/$A$ and equal charge radii. The same can be expected for $^{42}$Ti due to the difference in BE/$A$ to behave differently than the other semi-magic stable $^{50}$Ti isotope. The future measurement of spectroscopy as well as charge radii in $^{42}$Ti would be interesting in this regard. This is simply because binding of nucleons reflected in binding energy can govern the overall nuclear size and the charge radii. More tightly bound the nucleons are, lesser is the charge radius. Similarity in the charge radii of $^{40,48}$Ca isotopes is not only due to the doubly-magic configurations involved but also due to their similar binding energies which is not the case for the Ar and Ti isotopic chains.

\section{Conclusions}

The nuclear size, that is, isotopic shift for charge radii of the even-even $^{36-52}$Ca, $^{32-48}$Ar, and $^{44-50}$Ti has been studied in terms of the \mbox{IBM-CM}. A near-spherical nature of $^{36-52}$Ca isotopes is obtained, although core excitations are necessary to describe the yrare states. Bosonic wave functions are first fixed using the agreement of level energies and then used to compute the isotopic shift of charge radii in $^{36-52}$Ca isotopes. The $E0$ transitions connecting $0_2^+$ states to $0_1^+$ states are correlated with the isotopic shift of charge radii in Ca isotopes where $\rho^2(E0; 0_2^+ \rightarrow 0_1^+)$ data are also available for the comparison. With the obtained bosonic wave functions, a study on even-even Ar and Ti isotopes is also presented for the isotopic shift of charge radii. The BE/$A$ qualitatively explains the similarity of charge radius in $^{40}$Ca and $^{48}$Ca, and difference of charge radius in $^{38}$Ar and $^{46}$Ar, and $^{42}$Ti and $^{50}$Ti. It may be mentioned that the powerful connection between nuclei and neutron stars,
encoded in the nuclear equation of state,
in which the nuclear symmetry energy
(i.e., the cost of converting symmetric nuclear matter into pure neutron matter)
is used to set the basis of long-planned radius experiments
at JLab~\cite{adhikari2022,adhikari2021}. The current analysis provides an additional perspective for the same.

\begin{acknowledgements}
The author BM gratefully acknowledges the financial support from the HORIZON-MSCA-2023-PF-01 project, ISOON, under grant number 101150471. The funding from the Croatian Science Foundation and the \'Ecole Polytechnique F\'ed\'erale de Lausanne, under the project TTP-2018-07-3554 ``Exotic Nuclear Structure and Dynamics'', with funds of the Croatian-Swiss Research Programme is also acknowledged for enabling the initiation of this work. 
\end{acknowledgements}


\pagebreak

\end{document}